\documentclass{article}%
\usepackage{amsmath}%
\usepackage{amsfonts}%
\usepackage{amssymb}%
\usepackage{graphicx}
\providecommand{\U}[1]{\protect\rule{.1in}{.1in}}
\begin{document}

\title{On the significance of asperity models predictions of rough contact with
respect to recent alternative theories}
\author{M.Ciavarella}
\maketitle

\begin{abstract}
Recently, it has been shown that while asperity models show correctly
qualitative features of rough contact problems (linearity in area-load,
negative exponential dependence of load on separation which means also
linearity of stiffness with load), the exact value of the coefficients are not
precise for the idealized case of Gaussian distribution of heigths. This is
due to the intrinsic simplifications, neglecting asperity coalescence and
interaction effects. However, the issue of Gaussianity has not been proved or
experimentally verified in many cases, and here we show that, for example,
assuming a Weibull distribution of asperity heigths, the area-load linear
coefficient is not much affected, while the relationships load-separation and
therefore also stiffness-load do change largely, particularly when considering
bounded distributions of asperity heigths. 

It is suggested that Gaussianity of surfaces should be further tested in
experiments, before applying the most sophisticated rough contact models based
on the Gaussian assumption.

\end{abstract}

Addre\bigskip ss: Politecnico di BARI. V.le Gentile 182, 70125 Bari-Italy.

Email mciava@poliba.it, tel +390805962811 fax +390805962777

\bigskip

\section{Introduction}

\bigskip The GW model (Greenwood \&\ Williamson, 1966) has in the last 15-20
years been questioned and Persson's theories (Persson, 2001, 2007) have been
developed which, with significantly more elaborated modelling, capture the
correct behaviour of the problem of Gaussian rough surfaces with self-affine
spectrum. The two main criticisms are:

1) that linearity in area-load occurs only at asymptotically low loads in BGT
models, and otherwise a large spurious dependence on breadth parameter
$\alpha$ (as defined by Nayak, 1971) is predicted. In contrast, Persson's
theory is not apparently affected by $\alpha$ and shows a linear behavior
between contact area and load up to 10--15\% of the nominal contact area.

2) that the load-separation relationship should also be qualitatively
different, and even in the asymptotic range of high separation (Persson, 2007).

We illustrate these two deviations with respect to the simplest GW--McCool
model (McCool, 1992) which obtains area-separation and force-separation as
follows
\begin{align}
\frac{A_{c}\left(  t\right)  }{A_{0}} &  =\frac{\pi\sigma_{s}RD_{sum}}%
{\sqrt{2\pi}}I_{1}^{g}\left(  t\right)  \label{area}\\
\frac{F\left(  t\right)  }{A_{0}} &  =\frac{4}{3\sqrt{2\pi}}E^{\ast}\left(
R\sigma^{3}\right)  ^{1/2}D_{sum}I_{3/2}^{g}\left(  t\right)  \label{load}%
\end{align}
where $I_{n}^{g}\left(  t\right)  =\int_{t}^{\infty}d\xi\left(  \xi-t\right)
^{n}\exp\left(  -\xi^{2}/2\right)  $, and $\sigma$ is rms amplitude of summit
heights, $t=s/\sigma$ is dimensionless separation, $R$ their radii, $D_{sum}$
the density of summits, and $E^{\ast}$ the plain strain elastic modulus of the
contacting materials. From random process theory (Nayak, 1971), we can take%

\begin{equation}
D_{sum}=\frac{1}{6\pi\sqrt{3}}\frac{m_{4}}{m_{2}}\quad;\quad\frac{1}{R}%
=\frac{8}{3}\sqrt{\frac{m_{4}}{\pi}}\quad;\quad\sigma=\sqrt{m_{0}}\left(
1-\frac{0.9}{\alpha}\right)
\end{equation}
and hence $R\sigma D_{sum}=\frac{1}{48}\sqrt{\frac{3}{\pi}\left(
\alpha-0.9\right)  }$, where $m_{0},m_{2},m_{4}$ are the moments of the PSD,
or else the variance of surface heights, slopes and curvatures.

Further, the slope in the area-load relationship is found easily to be%
\begin{equation}
\frac{A_{c}\left(  t\right)  \Omega}{F\left(  t\right)  }=\frac{3^{3/2}\pi
}{8\sqrt{2}}\left(  \sqrt{\frac{1}{\alpha}}\right)  ^{1/2}\frac{I_{1}%
^{g}\left(  t\right)  }{I_{3/2}^{g}\left(  t\right)  }\simeq\frac{3^{3/2}\pi
}{8\sqrt{2}}\left(  \sqrt{\frac{1}{\alpha}}\right)  ^{1/2}\left(
1+0.148t\right)
\end{equation}
where $\Omega=E^{\ast}\sqrt{m_{2}/\pi}$, showing dependence on bandwidth. This
trend is in agreement with all other asperity contact theories at finite
separations, disregarding the fact that at very large separations, the more
refined ones predict linearity (Carbone \&\ Bottiglione, 2008). However, the
change with $\alpha^{-1/4}$ is largely erroneous.\ It is clear that this
factor, being possibly very large for very low fractal dimensions, may result
in a slope change of a factor $1000^{-1/4}=\allowbreak0.178$, i.e. almost an
order of magnitude in the quite extreme cases.

Regarding the force-separation law (and its derivative, the stiffness),
dividing this for the load itself, gives for the GW-McCool model (notice that
we omit the dependence on the rms amplitude $\sigma$ as we differentiate with
respect to $t$)
\begin{equation}
-\frac{\partial}{\partial t}\frac{F\left(  t\right)  }{A_{0}}/\frac{F\left(
t\right)  }{A_{0}}=-I_{3/2}^{g\prime}\left(  t\right)  /I_{3/2}^{g}\left(
t\right)  \simeq0.92\left(  1+t\right)
\end{equation}
giving for reasonable dimensionless separations $t=1,3$ a value between 2 and 3.5.

BGT models do not change this result significantly, as the load-separation is
not affected, contrary to the area-load, significantly by bandwidth, so that
we can take the asymptotic expression at high separations $\frac{F\left(
t\right)  }{A_{0}}\sim t^{-1}\exp\left(  -\frac{t^{2}}{2}\right)  $, which
differentiated gives
\begin{equation}
-\frac{\partial}{\partial t}\frac{F\left(  t\right)  }{A_{0}}/\frac{F\left(
t\right)  }{A_{0}}\sim\frac{\left(  t^{-2}+1\right)  \exp\left(  -\frac{t^{2}%
}{2}\right)  }{t^{-1}\exp\left(  -\frac{t^{2}}{2}\right)  }\sim t
\end{equation}
giving again for reasonable dimensionless separations $t=1,3$ a value between
1 and 3. Obviously one can fit a single value for the slope, finding an
average value.

Persson (2007) derives from elastic energy concepts and with extensive
derivations and correcting factors, a force-separation law $\frac{F\left(
t\right)  }{A_{0}}\sim\exp\left(  -\frac{t}{\beta}\right)  $, where $\beta$ is
a corrective factor of the order $0.5-1$ (higher values being for low fractal
dimensions), which however does not depend on separation and hence is a fixed
value for a given system. This results in
\begin{equation}
-\frac{\partial}{\partial t}\frac{F\left(  t\right)  }{A_{0}}/\frac{F\left(
t\right)  }{A_{0}}\sim\frac{1}{\beta}%
\end{equation}
Therefore Persson's $\beta=0.5-1$ corresponds to a slope between 1 and 2 which
however does not depend on separation this time, but only on the spectrum
geometrical characteristics. However, it is hard to say that results are
qualitatively different. Proportionality of stiffness with load is
approximately true for both theories, and perhaps more rigorously for
Persson's theory --- but this hasn't been shown to be so extensively valid in
experimental surfaces, with few good exceptions (Lorenz et al., 2010).
Numerical simulations of the dependence of stiffness on the applied squeezing
pressure show that the dependence on the large wavelengths component of
roughness induces large scatter in results and finite-size effects which
induce sublinear power-law scaling (Pastewka et al, 2013) which means higher
values for $\beta$ also depending on separation, but this is important in the
more general context of the reliability of a Gaussian random process
approximation. 

We shall return to the very basic point: are \textit{real surface Gaussian}?
Clearly, it did not seem so at the time of the original Greenwood and
Williamson (1966) paper, if one looks at Fig.6 where a mild steel had been
abraded and then put to slide against copper: so we are unlikely to make a
good approximation with a Gaussian if surfaces have undergone wear. We can
estimate what happens in wear processes from the closely related problem of
Chemical Mechanical\ Polishing in planarization of integrated circuits
(Vasilev et al, 2013, Borucki, 2002, Borucki \textit{et al.} 2004). Indeed,
polishing causes high asperities wearing faster than low asperities, and
Borucki derived, based on Archard law for wear and the GW model, an equation
similar to a differential Hamilton-Jacobi equation for the distribution of
asperity heights. High peaks in the tail of the distribution appear, the tail
becomes increasingly worn out and indeed even a delta function can result.
Stein et al (1996) show indeed experimental distribution of heights which
resemble inverted Weibull, i.e. where the tail is on the lower end of the
height distribution, and the right tail has been completely removed. 

Nayak (1971) is the classical reference for random process theory, yet he
himself writes that "\textit{it is clear that many surfaces are non-Gaussian;
but it is equally clear that many surface are Gaussian}". Central Limit
Theorem (CLT) is a weak justification for Gaussian height distribution, and
even numerical realization intended to be Gaussian in principle, sometimes
fail to do so. In fact, the longest wavelengths will dominate the height
distribution, and there are generally only very few of them. More in general,
Persson \textit{et al} (2005) has a number of surface height distributions
also from experiments, but it is unclear if they are more or less Gaussian,
and no statistical test is conducted. 

Here, we shall relax the assumption of Gaussian height distribution, while
maintaining the assumption of asperity behaviour, and in particular, we shall
consider the extreme case of a Weibull distribution of asperities approaching
a rigid plane from the bounded side of the distribution. It will be shown that
this leads to dramatic changes in the results, much more significant than the
differences between asperity models and Persson's theory, which have been
found in the specific, although important, case of Gaussian surfaces.

\section{Weibull asperity distribution}

According to the standard\ GW integration process as a function of asperity
height, we obtain number of asperities in contact, area and load, integrating
\begin{align}
n  & =N\frac{a}{\sigma_{s}}\int_{d_{0}}^{\infty}\left(  \frac{z_{s}}%
{\sigma_{s}}\right)  ^{a-1}\exp\left(  -\left(  \frac{z_{s}}{\sigma_{s}%
}\right)  ^{a}\right)  dz_{s}\\
\qquad A  & =N\pi R\frac{a}{\sigma_{s}}\int_{d_{0}}^{\infty}\left(
z_{s}-d_{0}\right)  \left(  \frac{z_{s}}{\sigma_{s}}\right)  ^{a-1}\exp\left(
-\left(  \frac{z_{s}}{\sigma_{s}}\right)  ^{a}\right)  dz_{s}\\
P  & =N\frac{a}{\sigma_{s}}\int_{d_{0}}^{\infty}\frac{4}{3}\frac{E}{R}\left(
Rd\right)  ^{3/2}\left(  \frac{z_{s}}{\sigma_{s}}\right)  ^{a-1}\exp\left(
-\left(  \frac{z_{s}}{\sigma_{s}}\right)  ^{a}\right)  dz_{s}%
\end{align}
where $N$ is number of asperities, $\sigma_{s}$ the amplitude parameter (which
does not correspond to the $\sigma$ introduced above as rms amplitude of
summit heights distribution), $R$ the radius of asperities.

Then, changing variable to $d+d_{0}=z_{s}$, normalizing the heights by
$\sigma_{s}$, the lower extreme of integration is $0$ and defining
\begin{equation}
I_{n}\left(  d_{0}^{\ast}\right)  =\int_{0}^{\infty}\left(  d^{\ast}\right)
^{n}\left(  d^{\ast}+d_{0}^{\ast}\right)  ^{a-1}\exp\left(  -\left(  d^{\ast
}+d_{0}^{\ast}\right)  ^{a}\right)  dd^{\ast}%
\end{equation}
we obtain%
\begin{align}
n\left(  d_{0}^{\ast}\right)    & =NaI_{0}\left(  d_{0}^{\ast}\right)  \\
\qquad A\left(  d_{0}^{\ast}\right)    & =\left(  NR\sigma_{s}\right)  \pi
aI_{1}\left(  d_{0}^{\ast}\right)  \\
P\left(  d_{0}^{\ast}\right)    & =\frac{4}{3}NaER^{1/2}\sigma_{s}%
^{3/2}I_{3/2}\left(  d_{0}^{\ast}\right)
\end{align}
and so the ratio between stiffness and load at a given separation is%
\[
\frac{\partial}{\partial d_{0}^{\ast}}P\left(  d_{0}^{\ast}\right)  /P\left(
d_{0}^{\ast}\right)  =I_{3/2}^{\prime}\left(  d_{0}^{\ast}\right)
/I_{3/2}\left(  d_{0}^{\ast}\right)
\]

We write non dimensional relationships,
\begin{align}
\qquad\frac{A\left(  d_{0}^{\ast}\right)  }{A_{0}^{g}}  & =\frac{a\sqrt{2\pi}%
}{\sqrt{\Gamma\left(  1+\frac{2}{a}\right)  -\Gamma\left(  1+\frac{1}%
{a}\right)  ^{2}}}I_{1}\left(  d_{0}^{\ast}\right)  \\
\frac{P\left(  d_{0}^{\ast}\right)  }{P_{0}^{g}}  & =\frac{a\sqrt{2\pi}%
}{\left(  \Gamma\left(  1+\frac{2}{a}\right)  -\Gamma\left(  1+\frac{1}%
{a}\right)  ^{2}\right)  ^{3/4}}I_{3/2}\left(  d_{0}^{\ast}\right)
\end{align}

Here, we have used the rms amplitude for a Weibull height distribution
$h_{rms}=\sigma_{s}\sqrt{\Gamma\left(  1+\frac{2}{a}\right)  -\Gamma\left(
1+\frac{1}{a}\right)  ^{2}}$), and we have defined the factors $P_{0}%
^{g}=\frac{4}{3}\frac{N}{\sqrt{2\pi}}ER^{1/2}h_{rms}^{3/2}$ and $A_{0}%
^{g}=\sqrt{\pi/2}\left(  NRh_{rms}\right)  $ to make easier comparison with
the Gauss distribution result. In the Gaussian case, we return essentially to
the McCool model, although here we are normalizing by the $P_{0}^{g},A_{0}%
^{g}$ factors..

\section{Results}

In Fig.1 we plot the standard results for the asperity model with Weibull
distribution, where it is clear that area-load is almost linear and depends
very little from the Weibull shape parameter. The bandwidth dependence cannot
appear from this plot, but it is not relevant. On the other hand, the
load-separation plots (Fig.1b) show very large change of the trend, and notice
that the Persson's result is intermediate between the exponential distribution
and the Rayleigh distribution ($a=1,2$ respectively). So nothing really new
here. 

More interesting is the case when the contact occurs on the tail of the
distribution, as in Fig.2.\ Here, area-load continues to be mostly unaffected
by the Weibull parameter, although it becomes more non-linear, but on the
contrary the load-separation becomes extremely non-linear even in the
log-linear scale. Clearly, this is mechanically simple to explain, as we have
a large number of asperities near the point of initial contact, so stiffness
increases very rapidly upon further indentation, and then tends to a very
"soft" value.

\begin{center}
$%
\begin{array}
[c]{cc}%
%TCIMACRO{\FRAME{itbpF}{3.659in}{2.2857in}{0in}{}{}{area-load-normal.eps}%
%width 3.659in;  height 2.2857in;  depth 0in;  original-width 3.6115in;
%original-height 2.2459in;  cropleft "0";  croptop "1";  cropright "1";
%cropbottom "0";  filename 'area-load-normal.eps';file-properties "XNPEU";}} }%
%BeginExpansion
{\includegraphics[
height=2.2857in,
width=3.659in
]%
{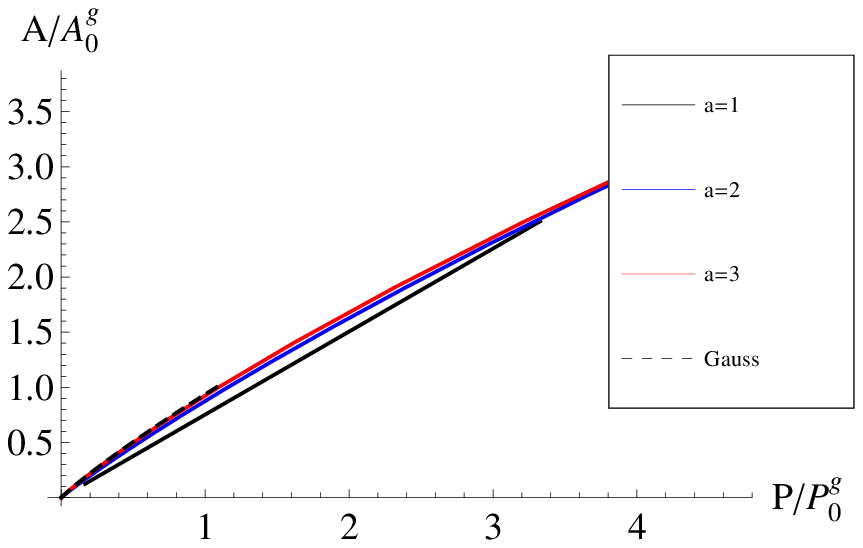}%
}
%EndExpansion
& (a)\\%
%TCIMACRO{\FRAME{itbpF}{3.659in}{2.1439in}{0in}{}{}{load-separation-normal.eps}%
%width 3.659in;  height 2.1439in;  depth 0in;  original-width 3.6115in;
%original-height 2.1049in;  cropleft "0";  croptop "1";  cropright "1";
%filename 'load-separation-normal.eps';file-properties "XNPEU";}} }%
%BeginExpansion
{\includegraphics[
height=2.1439in,
width=3.659in
]%
{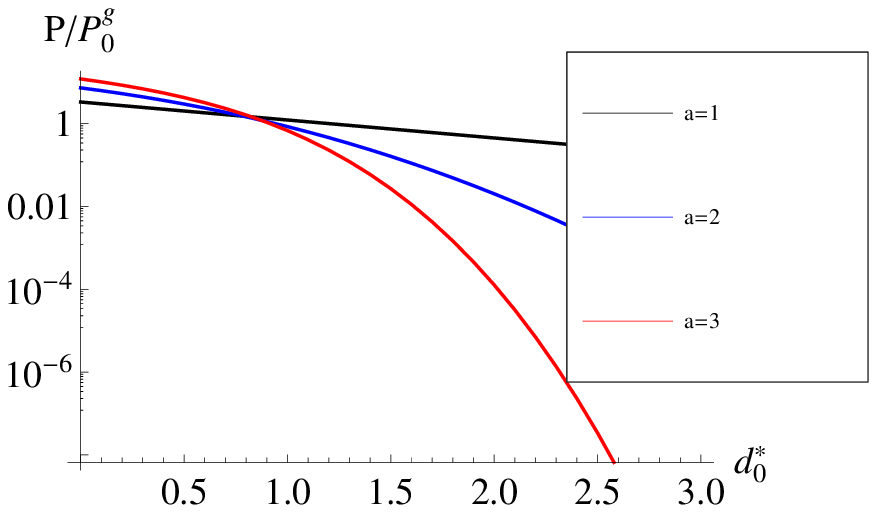}%
}
%EndExpansion
& (b)
\end{array}
$

Fig.1. Area-load (a), and load-separation, for the approach on the tail side
of the Weibull tail.\ Notice that separation is positive here as in standard
asperity models.

$%
\begin{array}
[c]{cc}%
%TCIMACRO{\FRAME{itbpF}{3.659in}{2.2857in}{0in}{}{}{area-load.eps}%
%width 3.659in;  height 2.2857in;  depth 0in;  original-width 3.6115in;
%original-height 2.2459in;  cropleft "0";  croptop "1";  cropright "1";
%cropbottom "0";  filename 'area-load.eps';file-properties "XNPEU";}} }%
%BeginExpansion
{\includegraphics[
height=2.2857in,
width=3.659in
]%
{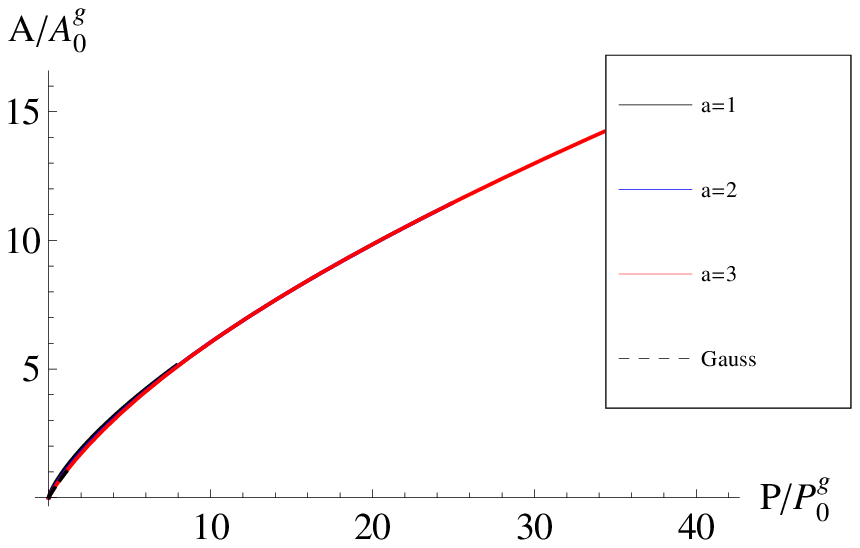}%
}
%EndExpansion
& (a)\\%
%TCIMACRO{\FRAME{itbpF}{3.659in}{2.1439in}{0in}{}{}{load-separation.eps}%
%width 3.659in;  height 2.1439in;  depth 0in;  original-width 3.6115in;
%original-height 2.1049in;  cropleft "0";  croptop "1";  cropright "1";
%cropbottom "0";  filename 'load-separation.eps';file-properties "XNPEU";}} }%
%BeginExpansion
{\includegraphics[
height=2.1439in,
width=3.659in
]%
{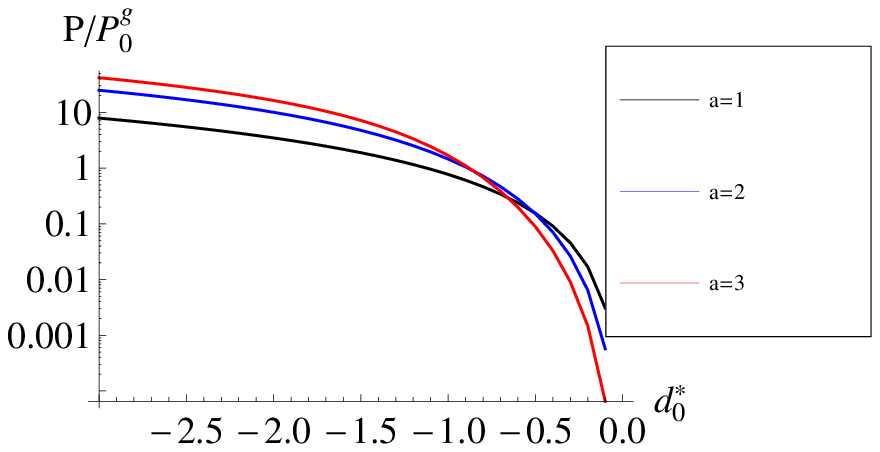}%
}
%EndExpansion
& (b)
\end{array}
$

Fig.2. Area-load (a), and load-separation, for the approach on the bounded
side of the Weibull tail.\ Notice that separation is negative in this case as
zero corresponds to just zero contact.

\bigskip
\end{center}

We further derive the results for stiffness in Fig.3, showing in Fig.3a the
results for the standard asperity model, showing the ratio stiffness to load
is actually lower than the Gaussian one (and hence quite similar to
the\ Persson's one), for low Weibull parameter, as observed already when
looking at the load-separation plots. Viceversa, when approach is on the
bounded side of the tail, the stiffness/load ratio approaches the value 1 only
at very large indentation, and otherwise it can increase substantially, being
in principle infinite at the point of initial contact. This suggests that the
detail of the height distribution matter, when considering stiffness.

\begin{center}
$%
\begin{array}
[c]{cc}%
%TCIMACRO{\FRAME{itbpF}{3.659in}{2.2857in}{0in}{}{}{stiffness-normal.eps}%
%width 3.659in;  height 2.2857in;  depth 0in;  original-width 3.6115in;
%original-height 2.2459in;  cropleft "0";  croptop "1";  cropright "1";
%cropbottom "0";  filename 'stiffness-normal.eps';file-properties "XNPEU";}} }%
%BeginExpansion
{\includegraphics[
height=2.2857in,
width=3.659in
]%
{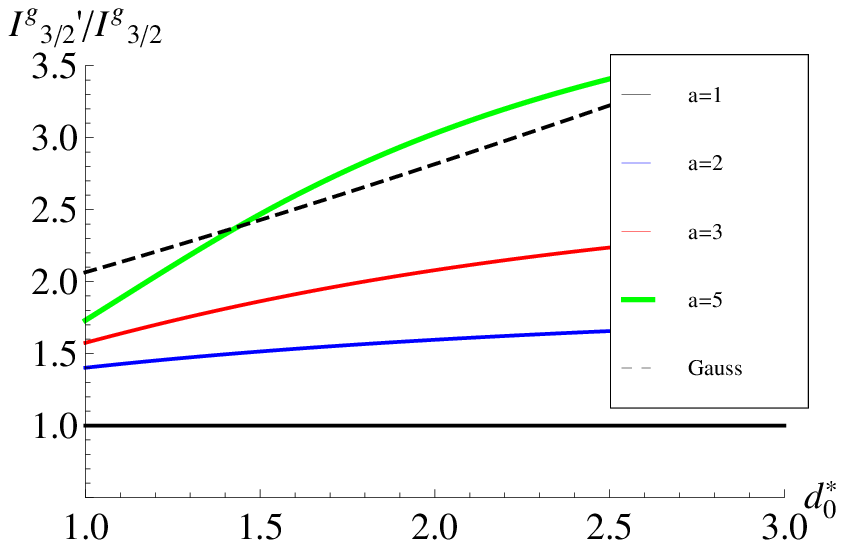}%
}
%EndExpansion
& (a)\\%
%TCIMACRO{\FRAME{itbpF}{3.659in}{2.271in}{0in}{}{}{stiffness.eps}%
%width 3.659in;  height 2.271in;  depth 0in;  original-width 3.6115in;
%original-height 2.2312in;  cropleft "0";  croptop "1";  cropright "1";
%cropbottom "0";  filename 'stiffness.eps';file-properties "XNPEU";}} }%
%BeginExpansion
{\includegraphics[
height=2.271in,
width=3.659in
]%
{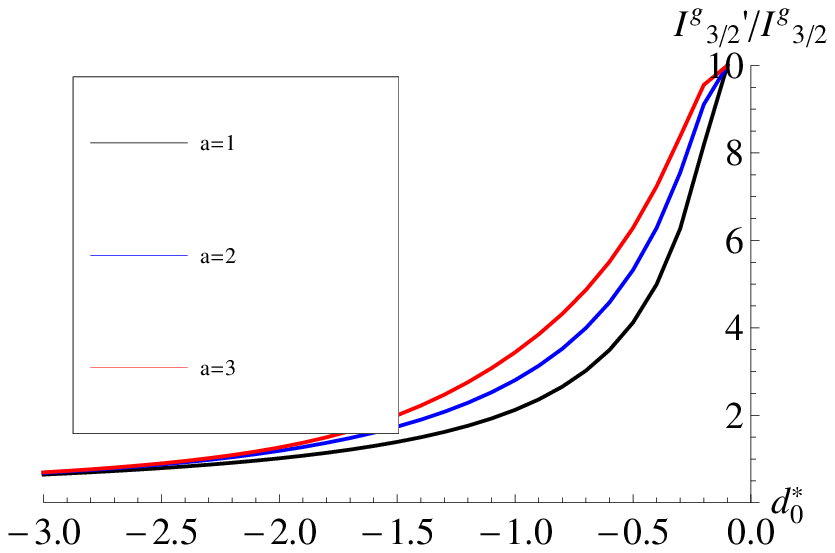}%
}
%EndExpansion
& (b)
\end{array}
$

Fig.3. Stiffness-load ratio, for the standard approach on the tail side of the
Weibull (a), or on the bounded side (b)

\bigskip

\end{center}

\section{Conclusion}

\bigskip We have shown that, while recent theories of contact mechanics like
that of Persson, have certainly improved our understanding of the Gaussian
surface problem, the assumption itself is quite strong, especially when it
comes to stiffness and relation load-separation, which largely depend on the
asperity height distribution.

\section{References}

Borucki, L. Mathematical modeling of polish rate decay in chemical-mechanical
polishing. J. Engng. Math. 43 (2002) 105--114

Borucki, L. J., Witelski, T., Please, C., Kramer, P. R., \& Schwendeman, D.
(2004). A theory of pad conditioning for chemical-mechanical polishing.
Journal of engineering mathematics, 50(1), 1-24.

Carbone, G., \& Bottiglione, F. (2008). Asperity contact theories: Do they
predict linearity between contact area and load?. Journal of the Mechanics and
Physics of Solids, 56(8), 2555-2572.

Greenwood, J.A., Williamson, J.B.P., 1966. Contact of nominally flat surfaces.
Proc. R. Soc. London A295, 300--319.

Lorenz, B., Carbone, G., \& Schulze, C. (2010). Average separation between a
rough surface and a rubber block: Comparison between theories and experiments.
Wear, 268(7), 984-990.

McCool, J. I. (1992). Non-Gaussian effects in microcontact. International
Journal of Machine Tools and Manufacture, 32(1), 115-123.

Nayak, P. R. 1971. Random process model of rough surfaces. Journal of
Tribology, 93(3), 398-407.

Pastewka, L., Prodanov, N., Lorenz, B., M\"{u}ser, M. H., Robbins, M. O., \&
Persson, B. N. (2013). Finite-size scaling in the interfacial stiffness of
rough elastic contacts. Physical Review E, 87(6), 062809.

Persson, B. N. 2001. Theory of rubber friction and contact mechanics. The
Journal of Chemical Physics, 115(8), 3840-3861.

Persson, B N J, Albohr, O, Tartaglino, U, Volokitin A I and Tosatti. E
\textit{\ }2005. On the nature of surface roughness with application to
contact mechanics, sealing, rubber friction and adhesion J. Phys.: Condens.
Matter 17 R1,\textit{\ http://arxiv.org/pdf/cond-mat/0502419.pdf)}

Persson, B. N. J. (2007). Relation between interfacial separation and load: a
general theory of contact mechanics. Physical review letters, 99(12), 125502.

Stein, D., Hetherington, D., Dugger, M., \& Stout, T. (1996). Optical
interferometry for surface measurements of CMP pads. Journal of Electronic
Materials, 25(10), 1623-1627.

Vasilev, B., Bott, S., Rzehak, R., \& Bartha, J. W. (2013). Pad roughness
evolution during break-in and its abrasion due to the pad-wafer contact in
oxide CMP. Microelectronic Engineering, 111, 21-28.
\end{document}